\documentclass[
  reprint,
  amsmath,amssymb,
  aps,
  prl,
  superscriptaddress,
  floatfix
]{revtex4-2}
\usepackage{float}
\usepackage{graphicx}
\usepackage{color}
\usepackage{xcolor}
\usepackage{dcolumn}
\usepackage{booktabs}
\usepackage{makecell}
\usepackage{bm}
\usepackage{capt-of}

\usepackage[colorlinks=true,
            linkcolor=blue,
            citecolor=blue,
            urlcolor=blue]{hyperref}

\makeatletter
\def\@seccntformat#1{\csname the#1\endcsname.\ }

\renewcommand\section{\@startsection{section}{1}{0pt}%
  {10pt plus 2pt minus 2pt}% space above
  {6pt}% space below
  {\normalfont\bfseries\centering}}

\renewcommand\subsection{\@startsection{subsection}{2}{0pt}%
  {8pt plus 2pt minus 2pt}% space above
  {4pt}% space below
  {\normalfont\normalsize\bfseries\centering}}
\makeatother

\begin{document}

% ---------------- TITLE + AUTHORS ----------------
\title{In-Line Fiber-Integrated Photon-Pair Generation from van der Waals Crystals}
\date{\today}

\author{Mayank Joshi}
\affiliation{Quantum Innovation Centre (Q.InC), Agency for Science, Technology and Research (A*STAR), Singapore 138635, Singapore}
\affiliation{Department of Quantum Science and Technology, Research School of Physics and Engineering, Australian National University (ANU), Canberra, ACT 2601, Australia}

\author{Tanumoy Pramanik}
\affiliation{Quantum Innovation Centre (Q.InC), Agency for Science, Technology and Research (A*STAR), Singapore 138635, Singapore}

\author{Mengting Jiang}
\affiliation{Institute of Materials Research and Engineering (IMRE), A*STAR, Singapore 138634, Singapore}

\author{Yu Xing}
\affiliation{Quantum Innovation Centre (Q.InC), Agency for Science, Technology and Research (A*STAR), Singapore 138635, Singapore}

\author{Zhaogang Dong}
\affiliation{Quantum Innovation Centre (Q.InC), Agency for Science, Technology and Research (A*STAR), Singapore 138635, Singapore}
\affiliation{Science, Mathematics, and Technology (SMT), Singapore University of Technology and Design, 8 Somapah Road, 487372, Singapore}

\author{Yuerui Lu}
\affiliation{School of Engineering, College of System and Society, ANU, Canberra, ACT, 2601, Australia}

\author{Jie Zhao}
\affiliation{Department of Quantum Science and Technology, Research School of Physics and Engineering, Australian National University (ANU), Canberra, ACT 2601, Australia}

\author{Ping Koy Lam}
\affiliation{Quantum Innovation Centre (Q.InC), Agency for Science, Technology and Research (A*STAR), Singapore 138635, Singapore}
\affiliation{Institute of Materials Research and Engineering (IMRE), A*STAR, Singapore 138634, Singapore}
\affiliation{Centre for Quantum Technologies (CQT), National University of Singapore (NUS), Singapore 117543, Singapore}

\author{Syed M. Assad}
\affiliation{Quantum Innovation Centre (Q.InC), Agency for Science, Technology and Research (A*STAR), Singapore 138635, Singapore}
\affiliation{Institute of Materials Research and Engineering (IMRE), A*STAR, Singapore 138634, Singapore}
\affiliation{Centre for Quantum Technologies (CQT), National University of Singapore (NUS), Singapore 117543, Singapore}

\author{Xuezhi Ma}
\email{ma\_xuezhi@a-star.edu.sg}
\affiliation{Quantum Innovation Centre (Q.InC), Agency for Science, Technology and Research (A*STAR), Singapore 138635, Singapore}
\affiliation{Institute of Materials Research and Engineering (IMRE), A*STAR, Singapore 138634, Singapore}
\affiliation{Centre for Quantum Technologies (CQT), National University of Singapore (NUS), Singapore 117543, Singapore}

\author{In Cheol Seo}
\email{incheol\_seo@a-star.edu.sg}
\affiliation{Quantum Innovation Centre (Q.InC), Agency for Science, Technology and Research (A*STAR), Singapore 138635, Singapore}
\affiliation{National Metrology Centre (NMC), Agency for Science, Technology and Research (A*STAR), Singapore 637145, Singapore}

\author{Young-Wook Cho}
\email{cho\_youngwook@a-star.edu.sg}
\affiliation{Quantum Innovation Centre (Q.InC), Agency for Science, Technology and Research (A*STAR), Singapore 138635, Singapore}
\affiliation{Institute of Materials Research and Engineering (IMRE), A*STAR, Singapore 138634, Singapore}
\affiliation{Centre for Quantum Technologies (CQT), National University of Singapore (NUS), Singapore 117543, Singapore}

% ---------------- ABSTRACT ----------------
\begin{abstract}

Miniaturized quantum light sources that operate directly in optical fibers are an attractive platform for optical quantum technologies. However, most miniaturized spontaneous parametric down-conversion (SPDC) sources still rely on objective-lens-based free-space pumping and collection, which limits compactness, robustness, and direct compatibility with fiber-based systems. Here we demonstrate a lens-free in-line SPDC photon-pair source by integrating a van der Waals NbOI$_2$ flake directly onto the end facet of an optical fiber. In this configuration, the generated photon-pairs are efficiently collected into optical fibers, eliminating the need for bulk free-space collection optics. Despite the limited numerical aperture of the single-mode fiber, efficient photon-pair collection with high purity, characterized by a coincidence-to-accidental ratio of up to $\sim$4600, is achieved in an ultracompact configuration. These results establish van der Waals ferroelectric materials as a promising platform for fiber-integrated quantum light sources and provide a pathway toward compact, alignment-free quantum photonic devices. 
\end{abstract}

\maketitle

% ==================================================
%  INTRODUCTION
% ==================================================
%\section{Introduction}
%\hspace{-.35cm}
\noindent

%%%%%%%%%%%%%%%%%%%%%%%%%%%%%%%%%%%%%%%%%%%%%%%%%%%%%
{\large \textbf{INTRODUCTION}}
%%%%%%%%%%%%%%%%%%%%%%%%%%%%%%%%%%%%%%%%%%%%%%%%%%%%%

Efficient and reliable sources of single-mode fiber-coupled photon-pairs are essential for photonic quantum technologies, including quantum computing \cite{l3}, networking \cite{l2,M1,M2}, sensing and quantum metrology \cite{M3,M4,M5,M6,M7,M8,Hong2021,Hong2022}. Spontaneous parametric down-conversion (SPDC) \cite{ll} in a nonlinear optical crystal is the most established and widely used mechanism for generating entangled photons \cite{M9,M10,M12,M13,M14,M15}. However, conventional SPDC implementations are typically based on bulk nonlinear optical crystals and free-space optical setups \cite{Choi2020,Cho2019}, which introduce alignment complexity and limit compact integration with optical fiber platforms. 

%. Furthermore, since conventional SPDC setups often rely on bulky free-space optics, which limits the direct integration of photon-pair sources with optical fibers.
%In these bulk crystals, the efficiency of photon-pair generation is strictly governed by rigid phase-matching conditions, where inherent material dispersion significantly constrains the operational frequency ranges~\cite{15, k1}. 

Emerging van der Waals (vdW) crystals have attracted intense research interest as a novel material platform for SPDC sources due to their giant non-linear susceptibility, atomic-scale thickness and the potential for scalable integration \cite{ref-transfer,aletheia2025nonlinear,l28}. More importantly, unlike bulk materials, the ultrathin nature of vdW crystals significantly relaxes the strict longitudinal phase-matching condition, enabling broader bandwidth photon-pair generation~\cite{M14,M15}. A variety of vdW crystals, including $\mathrm{NbOX_2}$ (X = Cl, Br, or I)  \cite{l5,bai,Br,l22}, 3R-MoS$_2$   \cite{l6,trovatello2025quasi}, 3R-WS$_2$  \cite{feng24}, r-BN  \cite{l9}, have been explored for SPDC process. Among them, layered niobium oxide diiodide ($\mathrm{NbOI_2}$) emerges as a particularly promising candidate due to its exceptionally large second-order nonlinear susceptibility ${\chi}^{(2)}$ of up to $\sim1000$ pm$/$V~\cite{l22}. Moreover, the vdW stacking without strict lattice matching provide a robust route for engineering SPDC sources  \cite{l11,l12,l13,l14}. For example, tailoring the stacking sequence and relative crystal orientation, quasi-phase matching \cite{l8} can be achieved for efficient photon-pair generation, as well as polarization entanglement generation and manipulation with high fidelity \cite{joshi2026air}.

%%%%%%%%%%%%%%%%%%%%%%%%%%%%%%
% Figure 1
%%%%%%%%%%%%%%%%%%%%%%%%%%%%%%
\begin{figure*}[tp]
\centering
\includegraphics[width= 6.5 in]{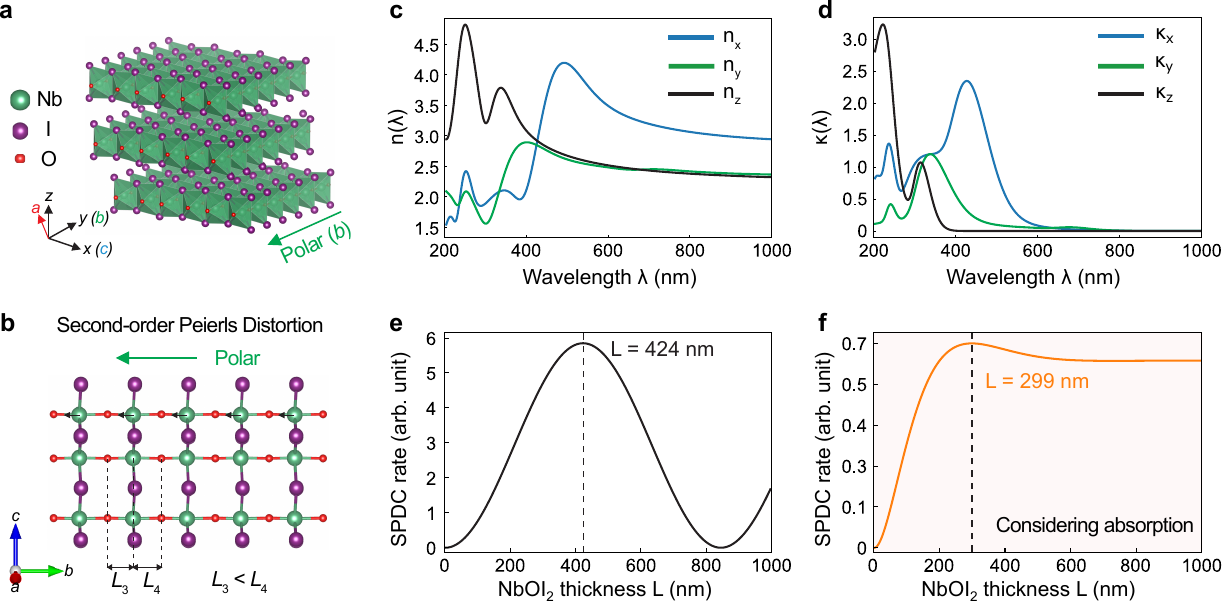}
\caption{
NbOI$_2$ properties for SPDC process.
\textbf{(a)} Crystal structure of layered NbOI$_2$ and definition of the laboratory axes ($x$,$y$,$z$) used in the measurements and their relation to the crystallographic axes ($a$,$b$,$c$). \textbf{(b)} Schematic illustration of the second-order Peierls distortion, in which Nb atoms displace along either the $+b$ or $-b$ direction, breaking inversion symmetry and giving rise to a nonzero ${\chi}^{(2)}$.
\textbf{(c)} Wavelength-dependent refractive index $n(\lambda)$ and
\textbf{(d)} extinction coefficient $\kappa(\lambda)$ along the principal axes, obtained from spectroscopic ellipsometry.
Calculated SPDC generation rate $R_{\mathrm{CC}}$ as a function of crystal thickness $L$: \textbf{(e)} without considering material absorption and \textbf{(f)} including absorption based on the complex refractive index. See the main text for further details.
}
\label{fig1}
\end{figure*}
%%%%%%%%%%%%%%%%%%%%%%%%%%%%%%

%Despite these advances, most demonstrations to date have been limited to free-space configurations with multi- mode fiber collection[29], which constraints spatial mode purity and ultimately limits further quantum state ma- nipulation via quantum interference. The ultrathin na- ture of vdW nonlinear crystals, however, enables non- linear interactions to occur directly at guide-wave inter- faces, including optical fiber facets. 
Despite these material advances, most demonstrations of 2D materials based ultrathin SPDC sources to date have been limited to free-space configurations with multi-mode fiber collection \cite{l5,feng24,l9,l20}. In free-space setups, efficiently coupling the emitted photons into fibers remains challenging due to spatial mode mismatch and alignment instability. To address these coupling issues, multimode fibers are often used, however, this reduces spatial mode purity and consequently limit further quantum state manipulation via subsequent quantum interference. The ultrathin nature of vdW nonlinear crystals, however, enables nonlinear interactions to occur directly at photonic interfaces, such as optical fiber facets, opening a pathway toward efficient single-mode fiber–coupled SPDC sources required for high-visibility quantum interference.

% inherently suffers from several spatial mode mismatches, reflection losses, and alignment instability. To overcome theses coupling issues, experiments often resort to multi-mode fibers, which, however, constraint spatial mode purity and ultimately limit further quantum state manipulation via quantum interference. The ultrathin nature of vdW nonlinear crystals, however, enables nonlinear interactions to occur directly at guide-wave interfaces, including optical fiber facets.

In this work, we integrate a vdW nonlinear crystal, $\mathrm{NbOI_2}$, onto the end facet of a single-mode fiber, realizing an in-line, fiber-coupled SPDC photon-pair source. In our configuration, the pump laser is delivered through an optical fiber and interacts with the $\mathrm{NbOI_2}$ crystal at the fiber facet, while the generate photon-pairs are directly coupled into a single-mode fiber. This fully fiber-integrated configuration eliminates the need for free-space alignment and bulk collection optics. Furthermore, we investigate several configurations of fiber-integrated devices and compare their photon-pair generation performance.

\par\vspace{1.2em}
%%%%%%%%%%%%%%%%%%%%%%%%%%%%%%%%%%%%%%%%%%%%%%%%%%%%%
{\large \textbf{RESULTS}}
%%%%%%%%%%%%%%%%%%%%%%%%%%%%%%%%%%%%%%%%%%%%%%%%%%%%%

%%%%%%%%%%%%%%%%%%%%%%%%%%%%%%%%%%
\textbf{SPDC from ultrathin materials} 
%%%%%%%%%%%%%%%%%%%%%%%%%%%%%%%%%%

%NbOI$_2$ has recently attracted considerable interests as nonlinear optical materials. Theoretically, strong excitation resonances in these crystals have been expected to enable efficient ultrathin SPDC sources. Experimentally, SPDC photon-pair generation has also recently been demonstrated in ultrathin NbOI$_2$ crystals. In our work, we employ NbOI$_2$ as the nonlinear medium for SPDC generation. 

%%%%%%%%%%%%%%%%%%%%%%%%%%%%%%
% Figure 2
%%%%%%%%%%%%%%%%%%%%%%%%%%%%%%
\begin{figure*}[tp]
\centering
\includegraphics[width= 6.5 in]{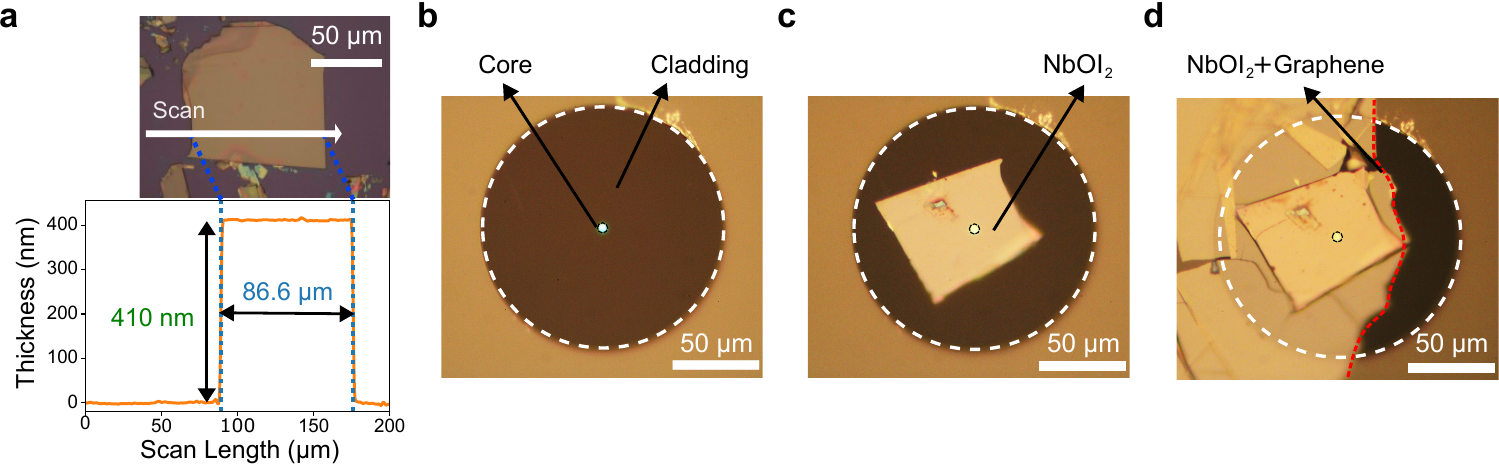}
\caption{
Preparation of the fiber-integrated NbOI$_2$ device. 
\textbf{(a)} Surface profile of the NbOI$_2$ flake used in this work, showing a thickness of $\sim$410~nm. Inset shows the optical image of the flake and the scan direction used for the measurement.
\textbf{(b)} Optical microscope image of the fiber facet showing the core and cladding. 
\textbf{(c)} NbOI$_2$ flake transferred onto the fiber core. Note that the original flake, shown in the inset of \textbf{(a)}, we split into two pieces before transfer. 
\textbf{(d)} NbOI$_2$ on the fiber facet is encapsulated by graphene layers with $\sim$35nm thickness.
}
\label{fig2}
\end{figure*}
%%%%%%%%%%%%%%%%%%%%%%%%%%%%%%

NbOI$_2$ is a vdW nonlinear crystal that combines strong optical anisotropy and a large  ${\chi}^{(2)}$, making it an attractive material platform for spontaneous parametric down-conversion (SPDC). Figure~\ref{fig1}(a) shows the layered crystal structure of NbOI$_2$ together with the definition of the coordinate axes used in the measurements. The laboratory coordinate system ($x$,$y$,$z$) is defined such that $x$ and $y$ axes lie within the layer plane, corresponding to crystallographic $c$ and $b$ axes, respectiely while the $z$ axis is perpendicular to the layer plane \cite{Xuan2024}. A key structural feature of NbOI$_2$ is its second-order Peierls distortion, illustrated in Fig.~\ref{fig1}(b), in which the Nb atoms are displaced along either the $+b$ or $-b$ direction. This distortion breaks inversion symmetry and gives rise to a nonzero second-order nonlinear susceptibility $\chi^{(2)}$, which underpins the SPDC process in this material\cite{l22}.

NbOI$_2$ is known to have highly anisotropic optical properties \cite{l22,Xuan2024}. To quantify the optical constants relevant to SPDC, we measured the complex refractive indices along the principal axes, $\tilde{n}_i(\lambda)=n_i(\lambda)+i\kappa_i(\lambda)$ with $i=x,y,z$, using spectroscopic ellipsometry. Here, $n_i(\lambda)$ and $\kappa_i(\lambda)$ denote the refractive index and extinction coefficient, respectively, with $\kappa_i(\lambda)$ characterizing optical absorption. The extracted wavelength-dependent refractive indices and extinction coefficients are shown in Fig.~\ref{fig1}(c) and (d), respectively. The measurements reveal strong optical anisotropy, which influences the polarization-dependent nonlinear optical responses relevant for SPDC. In particular, the extinction coefficient $\kappa_i (\lambda)$ becomes very small for wavelengths longer than 800 nm, indicating negligible absorption in this spectral range, which is favorable for SPDC photon generation. 

For a crystal of thickness $L$, the expected coincidence count rate of SPDC photons is given as \cite{l29}
\begin{equation}
R_{\mathrm{CC}} \propto A \left|\int^{L}_0 {\exp (i \Delta\tilde{k} z) dz} \right|^2 = A \left|\frac{e^{i\Delta\tilde{k} L}-1}{i\Delta\tilde{k}}\right|^2, 
\label{eq1}
\end{equation}
where the constant $A$ depends on the pump power and the second-order nonlinear susceptibility $\chi^{(2)}$, the complex phase mismatch is defined as $\Delta\tilde{k} = \tilde{k_p} - \tilde {k_s} - \tilde{k_i}$. Here, $\tilde{k}_j = 2\pi \tilde{n}_j$/$\lambda_j$ $(j=p,s,i)$ denotes the complex wavenumber of the pump, signal, and idler fields, respectively. $\lambda_p$, $\lambda_s$, and $\lambda_i$ denote the wavelength of the pump, signal, and idler photons. Note that we considered a collinear geometry in which the pump, signal, and idler photons propagate along the same direction. For a transparent crystal, Eq.~(\ref{eq1}) becomes  \cite{l29,l25}
\begin{equation}
R_{\mathrm{CC}} \propto A L^2
\left[
\mathrm{sinc}\!\left(\frac{\Delta k\, L}{2}\right)
\right]^2,
\label{eq2}
\end{equation}
where the complex phase mismatch $\Delta\tilde{k}$ becomes real, i.e. $\Delta\tilde{k} \rightarrow \Delta k$.

Figure~\ref{fig1}(d) shows the expected coincidence count rate as function of crystal thickness $L$ in the limit of weak absorption by using Eq.~(\ref{eq2}). The real valued phase mismatch $\Delta k$ is calculated using the refractive indices obtained from the ellipsometry measurements in Fig.~\ref{fig1}(b). Here, we select the pump wavelength of 405~nm and the signal/idler wavelengths of 810~nm. The pump, signal, and idler photons are all polarized along the $y$-axis. The calculation indicates that the SPDC photon-pair generation rate is maximized at the crystal coherence length $L_{\mathrm{coh}}=424$ nm and exhibits oscillatory behavior due to the non-zero phase mismatch. 

While most nonlinear optical crystals used for SPDC process are transparent, NbOI$_2$ exhibits substantial absorption at the pump wavelength (inferred from Fig.~\ref{fig1}(c)). As a result, the pump intensity decreases as it propagates through the crystal. To account for this effect, we also calculate the expected coincidence count rate by using Eq.~(\ref{eq1}). As shown in Fig.~\ref{fig1}(e), the oscillations are suppressed and the coincidence count rate gradually saturates with increasing crystal thickness. We found the maximum coincidence count rate at $L_{\mathrm{opt}}=299$~nm. 

%%%%%%%%%%%%%%%%%%%%%%%%%%%%%%
% Figure 3
%%%%%%%%%%%%%%%%%%%%%%%%%%%%%%
\begin{figure*}[tp]
\centering
\includegraphics[width= 6.5 in]{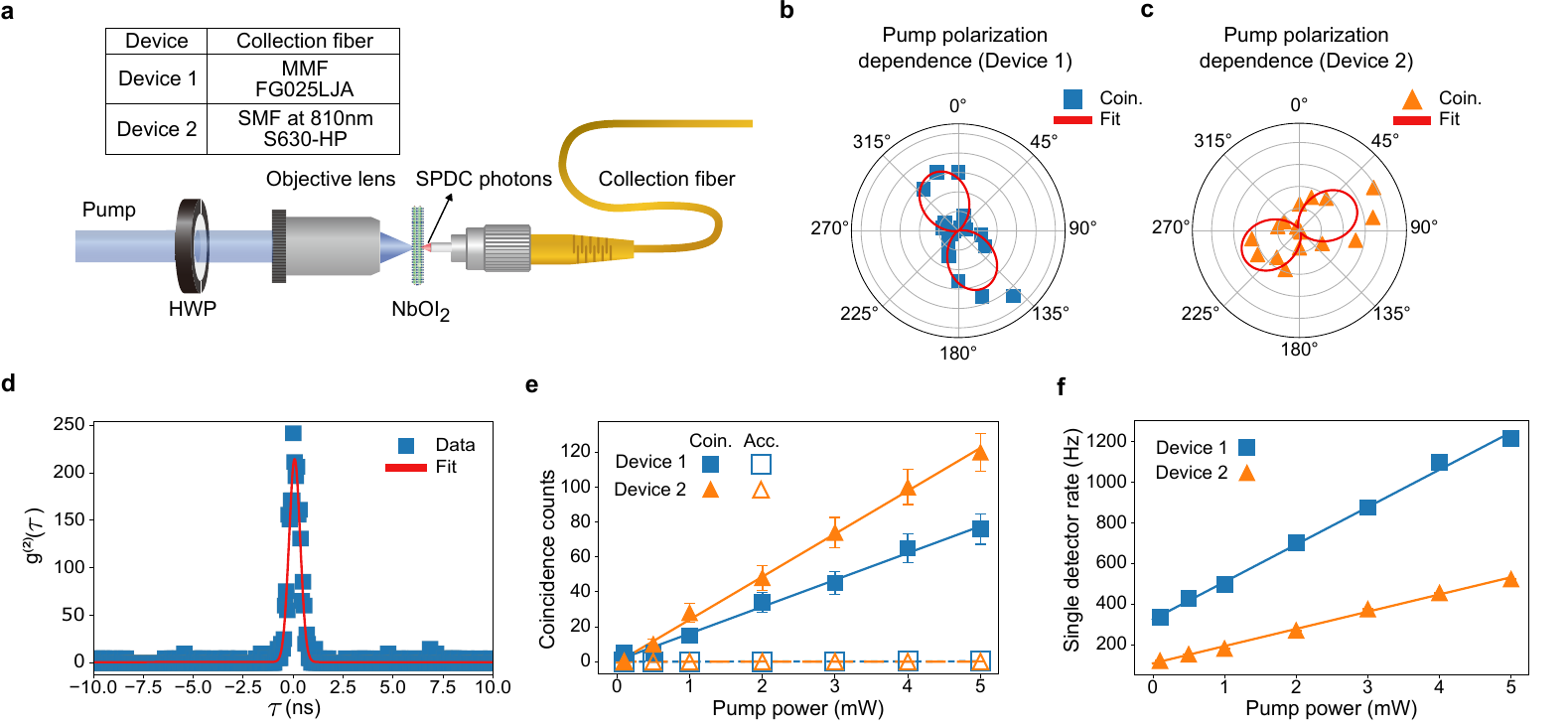}
\caption{
SPDC photon-pair generation from fiber-integrated NbOI$_2$ devices. 
\textbf{(a)} Experimental schematic for SPDC generation from an NbOI$_2$ flake placed on the fiber input facet. 
A continuous-wave 405 nm pump laser passes through a half-wave plate (HWP) for polarization control and is focused onto the NbOI$_2$ flake using a 10$\times$ objective lens.
The generated SPDC photons are directly coupled into the fiber and analyzed using a separate Hanbury Brown-Twiss (HBT) setup, see the main text for further details. 
\textbf{(b,c)} Pump polarization dependence of the SPDC coincidence counts for Device~1 (multimode fiber) and Device~2 (single-mode fiber), respectively.   
\textbf{(d)} Representative second-order correlation $g^{2}(\tau)$ measured for Device~1 as a function of the time delay $\tau$, measured with a bin width of 60~ps over an acquisition time of 30~min.  
\textbf{(e)} Coincidence counts and accidental counts as a function of pump power for Device~1 and Device~2. The coincidence window of 1.5~ns was used, and the coincidence counts were accumulated for 120~s. The error bars for the accidental counts largely overlap with lines and are therefore not clearly visible.
\textbf{(f)} Single-detector count rate as a function of pump power for Device~1 and Device~2. Error bars are smaller than the marker size and not visible. 
}
\label{fig3}
\end{figure*}
%%%%%%%%%%%%%%%%%%%%%%%%%%%%%%

%%%%%%%%%%%%%%%%%%%%%%%%%%%%%%%%%%
\textbf{Fiber-integrated NbOI$_2$.}
%%%%%%%%%%%%%%%%%%%%%%%%%%%%%%%%%%

Now, we prepare ultrathin NbOI$_2$ flakes with a thickness suitable for SPDC generation. The NbOI$_2$ flakes are mechanically exfoliated from bulk NbOI$_2$ crystals onto Si/SiO$_2$ substrate using the adhesive tape method. The thickness of the exfoliated flakes is first roughly estimated from their interference color under an optical microscope, and candidate flakes are selected accordingly. The thickness of the selected flakes is then measured more precisely using a surface profiler (KLA Tencor P16). The surface profile measurement, shown in Fig.~\ref{fig2}(a), confirms that the thickness of the NbOI$_2$ flake used in this work is measured to be $\sim 410$~nm. Although the calculated optimum occurs at $L_{opt}$=299~nm, the calculated SPDC rate is not strongly sensitive to the thickness for $L>$300~nm.

The selected NbOI$_2$ flake is subsequently transferred onto the end facet of a fiber using a polydimethylsiloxane (PDMS) stamp~\cite{l23}. The PDMS stamp is first used to pick up the flake from the Si/SiO$_2$ substrate and then carefully aligned with the fiber core (shown in Fig.~\ref{fig2}(b)) under an optical microscope. The flake is then released onto the fiber facet such that the NbOI$_2$ flake covers the fiber core region entirely, as shown in Fig.~\ref{fig2}(c). Due to the absence of dangling bonds in vdW materials, NbOI$_2$ can be easily stacked on diverse substrates without requiring stringent lattice matching~\cite{l17,l18,l19,ma2021engineering}, including fiber facets (we tested a multi-mode fiber (MMF) and a single-mode fiber (SMF)). After integrating the NbOI$_2$ flake onto the fiber facet, we encapsulated the flake with graphene layers as shown in Fig.~\ref{fig2}(d). Note that NbOI$_2$ undergoes the material degradation under ambient conditions with the optical property changes over time \cite{l21}. The graphene layers protect the NbOI$_2$ flake from environmental degradation such as humidity and oxidation, and also helps suppress pump-induced damage during the SPDC process \cite{joshi2026air}.

%%%%%%%%%%%%%%%%%%%%%%%%%%%%%%%%%%
\textbf{SPDC photon-pair generation in fiber-integrated NbOI$_2$ device.}
%%%%%%%%%%%%%%%%%%%%%%%%%%%%%%%%%%

To demonstrate SPDC photon-pair generation from the fiber-integrated NbOI$_2$ device, the crystal is pumped by a continuous-wave (CW) laser at 405~nm wavelength. As illustrated in Fig.~\ref{fig3}(a), the pump beam is focused onto the NbOI$_2$ flake using a $10\times$ objective lens. The linear polarization angle of the pump beam is controlled by a half-wave plate (HWP). The generated SPDC photons are directly collected by the fiber without any additional collection optics. The generated photon-pairs are then analyzed using a separate free-space Hanbury Brown-Twiss (HBT) setup (not shown in Fig.~\ref{fig3}(a)) \cite{HBT1956}. In the HBT setup, the fiber-coupled photon-pairs are collimated using another $10\times$ objective lens, and split by a 50:50 beam splitter, coupled into multi-mode fibers, and detected by two single-photon counting modules (SPCMs). After the beam splitter, bandpass filters centered at 810~nm with the full width half maximum (FWHM) bandwidth of 10~nm (Thorlabs FBH810-10) are placed in each arm to suppress residual pump and stray photons. Coincidence photon detection events between two SPCMs are then analyzed using a time-correlated single-photon counting (TCSPC), where a coincidence histogram of the photon arrival-time difference is constructed. 

We test two fiber-integrated devices: Device~1, integrated with a multimode fiber (FG025LJA, core diameter of $\sim$25 $\mu$m, NA of 0.10), and Device~2, integrated with a single-mode fiber (S630-HP, core diameter of $\sim$3.5 $\mu$m, single-mode field diameter of $\sim$4.2 $\mu$m at 630~nm, numerical aperture (NA) of 0.12). In order to have the identical thickness of NbOI$_2$ for both devices, we divided a single piece of NbOI$_2$ into two pieces. And, each piece is transferred to Device~1 and Device~2, respectively. 

First, we investigate the pump polarization dependence of the SPDC photon-pair generation. Figures~\ref{fig3}(b) and~\ref{fig3}(c) show the coincidence counts as a function of the pump polarization angle for Device~1 (multimode fiber) and Device~2 (single-mode fiber) respectively. In both cases, the coincidence counts exhibit a clear polarization dependence, which originates from the anisotropic nonlinear optical response of the NbOI$_2$ crystal. The SPDC generation efficiency reaches its maximum when the pump polarization is aligned along the crystal $y$-axis. In the following measurement, we set the pump polarization to be aligned along the $y$-axis.

%%%%%%%%%%%%%%%%%%%%%%%%%%%%%%
% Figure 4
%%%%%%%%%%%%%%%%%%%%%%%%%%%%%%
\begin{figure*}[tp]
\centering
\includegraphics[width= 6.5 in]{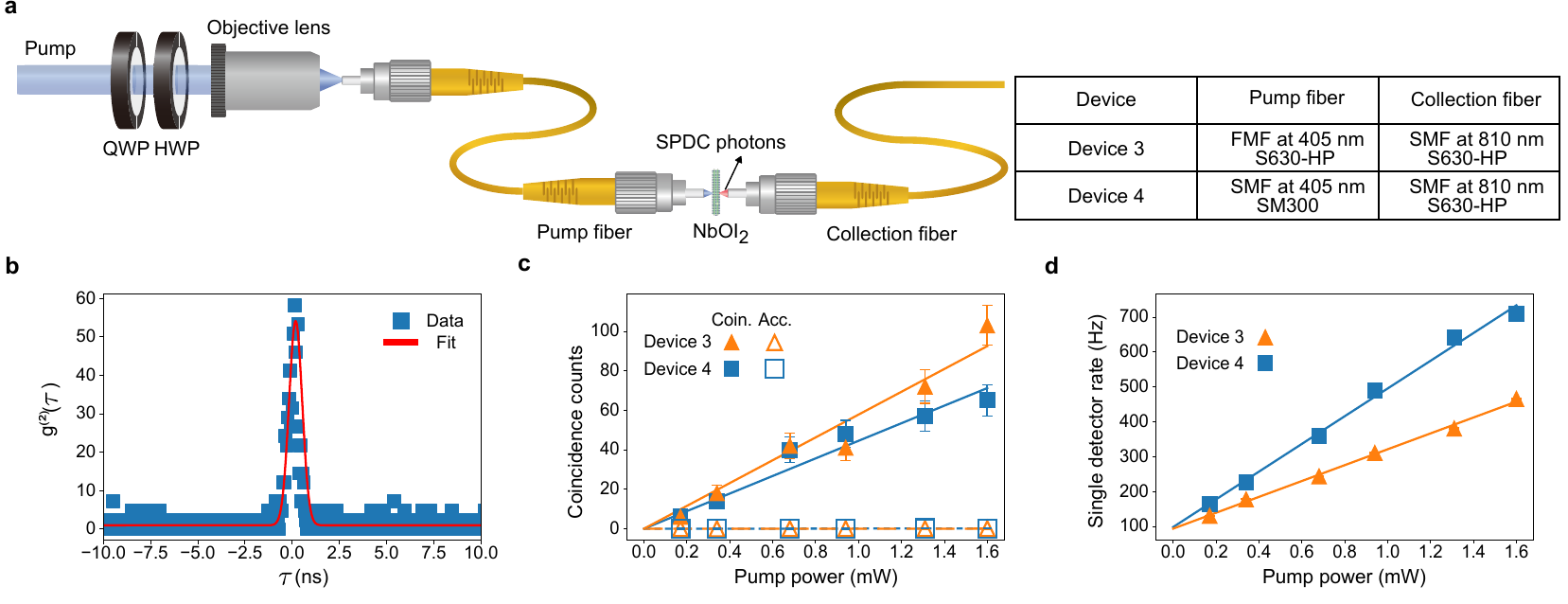}
\caption{ 
Fully fiber-integrated SPDC source based on NbOI$_2$.  
\textbf{(a)} Experimental configuration in which the pump laser is delivered through an optical fiber. The 405 nm pump passess through a quarter-wave plate (QWP) and a half-wave plate (HWP) for polarization control and is coupled to Fiber~1, which delivers the pump to the NbOI$_2$ crystal. Fiber~1 and Fiber~2 are connected using a standard FC-FC fiber adapter, with the NbOI$_2$ crystal positioned between the two fiber facets. The generated SPDC photons are collected by Fiber~2. 
\textbf{(b)} Representative second-order correlation $g^{2}(\tau)$ measured for Device~4 as a function of the time delay $\tau$, measured with a bin width of 60~ps over an acquisition time of 30~min.
\textbf{(c)} Coincidence counts and accidental counts as a function of pump power for Device~3 and Device~4. The coincidence window of 1.5~ns was used, and the coincidence counts were accumulated for 120~s. The error bars for the accidental counts largely overlap with lines and are therefore not clearly visible.  
\textbf{(d)} Single-detector count rate as a function of pump power for Device~3 and Device~4. Error bars are smaller than the marker size and not visible. 
}
\label{fig4}
\end{figure*}
%%%%%%%%%%%%%%%%%%%%%%%%%%%%%%

Figure~\ref{fig3}(d) shows a representative normalized second-order correlation function $g^{(2)}(\tau)$ obtained from TCSPC data accumulated for 30 minutes. This long acquisition time allowed us to clearly resolve the coincidence peak, from which we set our coincidence windows to 1.5~ns. Such time-resolved $g^{(2)}(\tau)$ measurements are not only useful for defining the coincidence window, but also verify the strong temporal quantum correlations between the SPDC photons. 

Using this coincidence window, we then measure the SPDC photon-pair generation rate as function of the pump power. Figures~\ref{fig3}(e) and~\ref{fig3}(f) show that both coincidence counts and single-detector count rate increase linearly with the pump power, as expected for the SPDC process. Note that the single-detector count rate, shown in Fig.~\ref{fig3}(f), is the geometric mean for both detectors. As shown in Fig.~\ref{fig3}(f), Device~1 (MMF) gives higher single-detector count rate than Device~2 (SMF), owing to its larger core diameter and the ability to collect photons over many spatial modes. The focused pump beam diameter $\sim$5.1~$\mu$m is much smaller than the multimode fiber core diameter $\sim$25~$\mu$m. Intuitively, one might therefore expect the coincidence counts to also be higher for Device~1. However, the experimental results show the opposite trend: as shown in Fig.~\ref{fig3}(e), the coincidence rate is higher for the single-mode fiber based device, Device~2. This indicates that Device~2 has a higher pair collection efficiency despite its lower single-count rates. Here, the pair collection efficiency \cite{Kwon08} is defined as 
%%%%%%%%%%%%%%%%%%%%%%%%%%%%%%%%%%%%%
\begin{equation}
\eta=R_{\mathrm{CC}}/\sqrt{R_{1} R_{2}},
\label{eff}
\end{equation}
%%%%%%%%%%%%%%%%%%%%%%%%%%%%%%%%%%%%%
where $R_{1}$ and $R_{2}$ denote the single-count rates measured at detector 1 and 2, respectively.

This result highlights that maximizing the single-photon collection efficiency does not necessarily maximize the pair collection efficiency of correlated photon-pairs \cite{Kwon08,Guerreiro2013,Schwaller2022}. SPDC generally produces photon-pairs over many transverse spatial modes. In our case, the ultrathin NbOI$_2$ crystal significantly relaxes the phase-matching constraint, leading to a broader angular emission spectrum and consequently a larger number of spatial modes \cite{Okoth2019}. As a result, a multimode fiber collects photons from many spatial modes, leading to higher single-count rates. In contrast, a single-mode fiber spatially filters the emission and selects a well-defined transverse mode. Since signal and idler photons are spatially correlated, this spatial filtering increases the pair collection efficiency \cite{Walborn2010}.

%%%%%%%%%%%%%%%%%%%%%%%%%%%%%%%%%%%%%%%%%
% Summary Table 
%%%%%%%%%%%%%%%%%%%%%%%%%%%%%%%%%%%%%%%%%
\begin{table*}[t]
\centering
\begin{tabular}{ccccccccc}
\toprule
Device   & \makecell{Pump\\ fiber} & \makecell{Collection\\ fiber}  & \makecell{$\omega_p$ \\ ($\mu$m)} & \makecell{$\omega_c$ \\ ($\mu$m)} & \makecell{Normalized\\ coincidence\\ rate\\ (Hz/mW)} & \makecell{Normalized\\single-detector\\ count rate\\ (Hz/mW)} & \makecell{Pair\\ collection\\efficiency\\ (\%)}  & CAR$_{\mathrm{max}}$\\
\midrule
Device 1 & free-space & MMF (NA=0.1) & $\sim$2.55 & $\sim$12.5 & $0.16 \pm 0.03$ & $209\pm0.78$ & $0.077 \pm 0.014$  & $365\pm65$ (at 4.0 mW) \\
Device 2 & free-space & SMF (NA=0.12) & $\sim$2.55 & $\sim$2.3  & $0.17 \pm 0.01$ & $82.5\pm 0.46$ & $0.21\pm 0.012$ & $2842\pm802$ (at 5.0 mW) \\
Device 3 & FMF& SMF (NA=0.12) & $\sim$1.60  & $\sim$2.3  & $0.43\pm 0.03$ & $216\pm0.78$ & $0.20\pm 0.014$ & $4643\pm2774$ (at 0.68 mW) \\
Device 4 & SMF& SMF (NA=0.12) & $\sim$1.25 & $\sim$2.3  & $0.38\pm 0.03$ & $393\pm1.1$ & $0.097\pm0.008$ & $2211\pm968$ (at 0.68 mW) \\
\bottomrule
\end{tabular}
\caption{Summary of the device configurations and SPDC performance of the four fiber-integrated devices. Note that normalized single-detector count rates are obtained after substracting detector dark counts. CAR$_\mathrm{max}$ indicates the maximum measured coincidence-to-accidental ratio (CAR) and the corresponding pump power.}
\label{table}
\end{table*}
%%%%%%%%%%%%%%%%%%%%%%%%%%%%%%%%%%%%%%

%%%%%%%%%%%%%%%%%%%%%%%%%%%%%%%%%%
\textbf{Fully fiber-integrated in-line NbOI$_2$ SPDC source.}
%%%%%%%%%%%%%%%%%%%%%%%%%%%%%%%%%%

We now realize a fully fiber-integrated in-line SPDC source, in which the pump laser is also guided through an optical fiber directly to the NbOI$_2$ crystal integrated on the fiber facet. For this configuration, Device~2 is reused and connected to an additional fiber via a standard FC-FC fiber adapter to deliver the pump laser. Two types of optical fibers were used to deliver the pump laser: S630-HP and SM300. The S630-HP, which is the same fiber used in Device~2, operates as a few-mode fiber (FMF) at the 405~nm pump wavelength, supporting two $\mathrm{LP}$ spatial mode groups ($\mathrm{LP}_{01}$ and $\mathrm{LP}_{11}$), corresponding to six vector modes. In contrast, the SM300 fiber remains single-mode at 405 nm. The SM300 fiber has a single-mode field diameter of $\sim$2.2~$\mu$m at 350~nm and NA of 0.13. The configuration using the S630-HP for the pump delivery is referred to as Device~3 (FMF), while the configuration using the SM300 fiber is referred to as Device~4 (SMF). 

Figure~\ref{fig4} shows the photon-pair generation from the fully fiber-integrated configuration, namely Device~3 and Device~4. The simplified experimental schematic is again illustrated in Fig.~\ref{fig4}(a) for clarity. The 405~nm pump laser is guided through an optical fiber and the pump polarization at the NbOI$_2$ crystal is controlled using a quarter-wave plate (QWP) and a half-wave plate (HWP) before coupling into the input fiber. Figure~\ref{fig4}(b-d) summarize the photon-pair generation results from the fully fiber-integrated devices (Device~3 and Device~4). During the measurements, the pump polarization was adjust to maximize the SPDC generation. A strong clear coincidence peak is observed in Fig.~\ref{fig4}(b), confirming again the generation of correlated photon-pairs.  

The coincidence counts and single-detector count rates as a function of pump power are shown in Fig.~\ref{fig4}(c) and~\ref{fig4}(d), respectively. Device~3 exhibits higher coincidence count rates than Device~4, but lower single-detector count rate. This indicates that Device~3 achieves a higher pair collection efficiency. The two devices differ only in the spatial mode of the pump beam defined by the pump fiber, while the collection configuration remains identical (In Device~3, the pump is delivered through a few-mode fiber, so the pump polarization may not be perfectly aligned with the crystal $y$-axis). However, this affects only the absolute pair generation rate and not the pair collection efficiency. This distinction is particularly important for multi-photon experiments, where the 2N-photon coincidence detection rate scales as $\eta^{N}$ \cite{Wang2016}, as well as for loophole-free Bell tests that requires high pair detection efficiency to close the fair-sampling loophole \cite{Shalm2015}. 

For this reason, considerable effort has been made to improve the pair collection efficiency of SPDC sources coupled to single-mode fibers. Efficient single-mode coupling requires spatial mode matching between the SPDC emission and the fiber fundamental mode, which can be achieved by optimizing the pump focusing condition and by using imaging optics to map the SPDC interaction region onto the fiber mode. Previous studies have shown that the optimal pump waist $\omega_p$ is comparable to the collection mode waist $\omega_c$, approximately satisfying $\omega_p \approx \omega_c/\sqrt{2}$ \cite{Ling2008}. This is consistent with our experimental observations. The pump beam waist at 405~nm is roughly estimated to be $\omega_p^3\sim$1.6~$\mu$m (for $\mathrm{LP}_{01}$ mode) for Device~3 and $\omega_{p}^4\sim$1.25~$\mu$m for Device~4. The collection mode waist at 810~nm determined by the collection fiber is estimated to be $\omega_c\sim$2.3~$\mu$m. Note that $\omega_p^3$ is close to the optimal value discussed above, i.e. $\omega_p^3 \approx \omega_c /\sqrt{2} =1.63~\mu$m.

%%%%%%%%%%%%%%%%%%%%%%%%%%%%%%%%%%
\textbf{Discussion}
%%%%%%%%%%%%%%%%%%%%%%%%%%%%%%%%%%

Table.~\ref{table} summarizes the key experimental configurations and performance metrics of the four fiber-integrated SPDC devices we have studied. Notably, Device~1, which employs multimode fiber collection, exhibits the lowest pair collection efficiency. This behavior is consistent with the fact that SPDC photons are emitted into multiple spatial modes, and multimode collection with limited NA increases the probability of detecting uncorrelated photons originating from different spatial modes of the SPDC emission. In free-space multimode fiber collection schemes, high-NA objective lenses therefore play an important role by capturing a larger fraction of the correlated photon-pairs emitted over a wide angular distribution. 

In contrast, devices employing single-mode fiber collection show significantly improved pair collection efficiencies up to 0.21\%, highlighting the role of spatial mode filtering in enhancing correlated photon detection. For single-mode fiber coupling, efficient coupling critically depends on proper spatial mode matching between the SPDC emission defined by the pump spatial mode, and the collection fiber mode, as discussed earlier. Device~3 represents the most ideal configuration among the devices studied and exhibits the highest normalized coincidence count rate as well as the largest coincidence-to-accidental ratio (CAR). However, its pair collection efficiency remains comparable to that of Device~2 within the experimental uncertainty. We attribute this to the intrinsic characteristics of the few-mode fiber (FMF) used for pump delivery and potential center-offset errors between the pump and collection fibers, which can degrade the spatial mode overlap and limit the achievable collection efficiency. It should also be noted that the overall brightness of photon-pairs collected into the single-mode fiber can be further improved by employing fibers with a higher NA without compromising the pair collection efficiency. 

\par\vspace{1.2em}
%%%%%%%%%%%%%%%%%%%%%%%%%%%%%%%%%%%%%%%%%%%%%%%%%%%%%
{\large \textbf{CONCLUSION}}
%%%%%%%%%%%%%%%%%%%%%%%%%%%%%%%%%%%%%%%%%%%%%%%%%%%%%

We have demonstrated SPDC photon-pair generation based on a vdW nonlinear crystal integrated directly on an optical fiber facet. Using NbOI$_2$ flakes, we realized a fiber-coupled SPDC source in which the generated photon-pairs are directly collected into optical fibers without relying on bulk free-space collection optics. Despite the limited numerical aperture (NA=0.12) of the single-mode fiber, we achieve a high pair collection efficiency of up to 0.21\% together with high-purity SPDC, reaching a coincidence-to-accidental ratio as high as $\mathrm{CAR_{max}}=4643$. To the best of our knowledge, the measured $\mathrm{CAR}$ of $\sim$4600 significantly exceeds previously reported values for vdW-based SPDC sources, such as the $\mathrm{CAR} >800 $ reported for 3R-WS$_2$ crystals \cite{feng24}.

To summarize, we have demonstrated a fully fiber-integrated in-line SPDC source in which both the pump laser and the generated photon-pairs are guided through optical fibers. This simple architecture eliminates the need for free-space alignment and provides improved stability. Such a fiber-integrated SPDC source provides a practical platform for future two-photon quantum interference experiments directly using optical fibers. 
$$
~~
$$

%%%%%%%%%%%%%%%%%%%%%%%%%%%%%%%%%%
\textbf{Acknowledgements}
%%%%%%%%%%%%%%%%%%%%%%%%%%%%%%%%%%

We thank Yoon-Ho Kim and Leonid A. Krivitsky for fruitful discussions. 

%%%%%%%%%%%%%%%%%%%%%%%%%%%%%%%%%%
\textbf{Author contributions}
%%%%%%%%%%%%%%%%%%%%%%%%%%%%%%%%%%

Y.-W.C., I.C.S., X.M., S.M.A. and J.Z. conceived and supervised the project. M.Jo. led the experiment and collected the data. All authors analyzed the data, discussed the results, and contributed to the manuscript.

%%%%%%%%%%%%%%%%%%%%%%%%%%%%%%%%%%
\textbf{Funding}
%%%%%%%%%%%%%%%%%%%%%%%%%%%%%%%%%%

This work was supported by the Agency for Science, Technology and Research (A*STAR) under the MTC YIRG Grants No. M23M7c0129 and No. M25N8c0139, and by the ASTAR Quantum Innovation Centre (Q.InC) SRTT core funding. This work was also supported by the National Research Foundation, Singapore, through the National Quantum Office hosted in A*STAR, under the Centre for Quantum Technologies Funding Initiative (S24Q2d0009). J.Z. acknowledges support from the Australian Research Council (ARC) Discovery Early Career Research Award (DECRA), Grant No. DE260101046.

%%%%%%%%%%%%%%%%%%%%%%%%%%%%%%%%%%
\textbf{Data availability}
%%%%%%%%%%%%%%%%%%%%%%%%%%%%%%%%%%

The data in this study are available from the corresponding author upon reasonable request.

%%%%%%%%%%%%%%%%%%%%%%%%%%%%%%%%%%
\textbf{Competing interests}
%%%%%%%%%%%%%%%%%%%%%%%%%%%%%%%%%%

The authors declare they have no competing interests.

\section{Reference}

\bibliographystyle{apsrev4-2}
\bibliography{ref_yw}

@article{ll,
  author = {A. Anwar and C. Perumangatt and F. Steinlechner and T. Jennewein and A. Ling},
  title   = {{Entangled photon-pair sources based on three-wave mixing in bulk crystals}},
  journal = {Review of Scientific Instruments},
  volume  = {92},
  number = {4},
  pages   = {041101},
  year    = {2021},
  note = {\url{https://doi.org/10.1063/5.0023103}}
}

@article{l2, 
  author = {I. Marcikic and H. De Riedmatten and W. Tittel and H. Zbinden and N. Gisin},
  title={{Long-distance teleportation of qubits at telecommunication wavelengths}},
  journal={Nature},
  volume={421},
  number={6922},
  pages={509--513},
  year={2003},
  publisher={Nature Publishing Group UK London},
  note = {\url{https://doi.org/10.1038/nature01376}}
}

@article{M1,
  title={{Quantum communication}},
  author = {N. Gisin and R. Thew},
  journal={Nature photonics},
  volume={1},
  number={3},
  pages={165--171},
  year={2007},
  publisher={Nature Publishing Group UK London},
  note = {\url {https://doi.org/10.1038/nphoton.2007.22}}
}

@article{M2,
  author = {F. Flamini and N. Spagnolo and F. Sciarrino},
  title = {{Photonic quantum information processing: a review}},
  journal = {Reports on Progress in Physics},
  volume = {82},
  number = {1},
  pages = {016001},
  year = {2018},
  publisher = {IOP Publishing},
  note={\url {https://doi.org/10.1088/1361-6633/aad5b2}}}

@article{M3,
  author = {Y. Shih},
  title   = {{Quantum imaging}},
  journal={IEEE Journal of Selected Topics in Quantum Electronics}, 
  volume  = {13},
  number={4},
  pages   = {1016--1030},
  year    = {2007},
  note={\url {https://doi.org/10.1109/JSTQE.2007.902724}}}

@article{M4,
  title={{Quantum imaging with undetected photons}},
  author = {G. B. Lemos and V. Borish and G. D. Cole and S. Ramelow and R. Lapkiewicz and A. Zeilinger},
  journal={Nature},
  volume={512},
  number={7515},
  pages={409--412},
  year={2014},
  publisher={Nature Publishing Group UK London},
  note={\url {https://doi.org/10.1038/nature13586}}
}

@article{M5,
  title={{Perspectives for applications of quantum imaging}},
  author={Gilaberte Basset, Marta and Setzpfandt, Frank and Steinlechner, Fabian and Beckert, Erik and Pertsch, Thomas and Gr{\"a}fe, Markus},
  journal={Laser \& Photonics Reviews},
  volume={13},
  number={10},
  pages={1900097},
  year={2019},
  publisher={Wiley Online Library},
  note={\url {https://doi.org/10.1002/lpor.201900097}}
}

@article{M6,
  author = {V. Giovannetti and S. Lloyd and L. Maccone},
  title= {{Advances in quantum metrology}},
  journal={Nature photonics},
  volume  = {5},
  number={4},
  pages= {222--229},
  year= {2011},
  note={\url {https://doi.org/10.1038/nphoton.2011.35}}
}

@article{M7,
  author = {C. L. Degen and F. Reinhard and P. Cappellaro},
  title   = {{Quantum sensing}},
  journal={Reviews of modern physics},
  volume  = {89},
  number={3},
  pages= {035002},
  year= {2017},
  note={\url {https://doi.org/10.1103/RevModPhys.89.035002}}
}

@article{M8,
  author = {L. Pezz\`e and A. Smerzi and M. K. Oberthaler and R. Schmied and P. Treutlein},
  title   = {{Quantum metrology with nonclassical states of atomic ensembles}},
  journal={Reviews of modern physics},
  volume  = {90},
  issue = {3},
  pages   = {035005},
  year    = {2018},
  note={\url {https://doi.org/10.1103/RevModPhys.90.035005}}
}

@article{M9,
  author = {{Horodecki, Ryszard and Horodecki, Pawe\l{} and Horodecki, Micha\l{} and Horodecki, Karol}},
  title   = {{Quantum entanglement}},
  journal={Reviews of modern physics},
  volume  = {81},
  issue = {2},
  pages   = {865--942},
  year    = {2009},
  note={\url {https://doi.org/10.1103/RevModPhys.81.865}}
}

@article{M10,
  author={Tanzilli, S{\'e}bastien and De Riedmatten, Hugues and Tittel, Wolfgang and Zbinden, Hugo and Baldi, Pascal and De Micheli, Marc and Ostrowsky, Daniel Barry and Gisin, Nicolas},
  title   = {{Highly efficient photon-pair source using periodically poled lithium niobate waveguide}},
  journal={Electronics Letters},
  volume  = {37},
  number={1},
  pages   = {26--28},
  year    = {2001},
  note={\url {https://doi.org/10.1049/el:20010009}}
}

@article{M12,
  author = {P. G. Kwiat and E. Waks and A. G. White and I. Appelbaum and P. H. Eberhard},
  title   = {{Ultrabright source of polarization-entangled photons}},
  journal={Physical Review A},
  volume  = {60},
  number={2},
  pages   = {R773--R776},
  year    = {1999},
  note={\url {https://doi.org/10.1103/PhysRevA.60.R773}}
}

@article{M13,
  author = {P. G. Kwiat and K. Mattle and H. Weinfurter and A. Zeilinger and A. V. Sergienko and Y. Shih},
  title   = {{New high-intensity source of polarization-entangled photon pairs}},
  journal={Physical Review Letters},
  volume  = {75},
  number={24},
  pages   = {4337--4341},
  year    = {1995},
  note={\url {https://doi.org/10.1103/PhysRevLett.75.4337}}
}

@article{M14,
  author={Okoth, Cameron and Kovlakov, E and B{\"o}nsel, F and Cavanna, Andrea and Straupe, S and Kulik, SP and Chekhova, MV},
  title   = {{Idealized Einstein-Podolsky-Rosen states from non-phase-matched parametric down-conversion.}},
  journal={Physical Review A},
  volume  = {101},
  number={1},
  pages   = {011801},
  year    = {2020},
  note={\url {https://doi.org/10.1103/PhysRevA.101.011801}}
}

@article{M15,
  author={Santiago-Cruz, Tom{\'a}s and Sultanov, Vitaliy and Zhang, Haizhong and Krivitsky, Leonid A and Chekhova, Maria V},
  title   = {{Entangled photons from subwavelength nonlinear films.}},
  journal={Optics Letters},
  volume  = {46},
  number={3},
  pages   = {653-656},
  year    = {2021},
  note={\url {https://doi.org/10.1364/OL.411176}}
}

@article{l3,
  author = {E. Knill and R. Laflamme and G. J. Milburn},
  title   = {{A scheme for efficient quantum computation with linear optics}},
  journal = {Nature},
  volume  = {409},
  number={6816},
  pages   = {46--52},
  year    = {2001},
  note={\url {https://doi.org/10.1038/35051009}}
}

@article{l5,
  author = {Q. Guo and X.-Z. Qi and L. Zhang and M. Gao and S. Hu and W. Zhou and W. Zang and X. Zhao and J. Wang and B. Yan and others},
  title   = {{Ultrathin quantum light source with van der Waals NbOCl$_2$ crystal}},
  journal = {Nature},
  volume  = {613},
  number={7942},
  pages   = {53--59},
  year    = {2023},
  note={\url {https://doi.org/10.1038/s41586-022-05393-7}}
}

@article{l6,
  author={Weissflog, Maximilian A and Fedotova, Anna and Tang, Yilin and Santos, Elkin A and Laudert, Benjamin and Shinde, Saniya and Abtahi, Fatemeh and Afsharnia, Mina and P{\'e}rez P{\'e}rez, Inmaculada and others},
  title   = {{A tunable transition metal dichalcogenide entangled photon-pair source}},
  journal={Nature Communications},
  volume  = {15},
  number={1},
  pages   = {7600},
  year    = {2024},
  note={\url {https://doi.org/10.1038/s41467-024-51843-3}}
}

@article{l8,
  author = {Y. Tang and K. Sripathy and H. Qin and Z. Lu and G. Guccione and J. Janousek and Y. Zhu and M. M. Hasan and Y. Iwasa and P. K. Lam and Y. Lu},
  title   = {Quasi-phase-matching enabled by van der Waals stacking},
  journal={{Nature Communications}},
  volume={15},
  number={1},
  pages={9979},
  year={2024},
  note={\url {https://doi.org/10.1038/s41467-024-53472-2}}
}

@article{l9,
  author = {H. Liang and T. Gu and Y. Lou and C. Yang and C. Ma and J. Qi and A. A. Bettiol and X. Wang},
  title = {{Tunable polarization entangled photon-pair source in rhombohedral boron nitride}},
  journal = {Science Advances},
  volume = {11},
  number = {4},
  pages = {eadt3710},
  year = {2025},
  note={\url {https://doi.org/10.1126/sciadv.adt3710}}
}

@article{l11,
  author = {A. K. Geim and I. V. Grigorieva},
  title={{Van der Waals heterostructures}},
  journal={Nature},
  volume={499},
  number={7459},
  pages={419--425},
  year={2013},
  note={\url {https://doi.org/10.1038/nature12385}}
}

@article{l12,
  author = {M.-Y. Li and C.-H. Chen and Y. Shi and L.-J. Li},
  title = {{Heterostructures based on two-dimensional layered materials and their potential applications}},
  journal = {Materials Today},
  volume = {19},
  number = {6},
  pages = {322-335},
  year = {2016},
  note={\url {https://doi.org/10.1016/j.mattod.2015.11.003}}
}

@article{l13,
  author = {H. Lim and S. I. Yoon and G. Kim and A.-R. Jang and H. S. Shin},
  title = {{Stacking of Two-Dimensional Materials in Lateral and Vertical Directions}},
  journal = {Chemistry of Materials},
  volume = {26},
  number = {17},
  pages = {4891-4903},
  year = {2014},
  note={\url {https://doi.org/10.1021/cm502170q}}
}

@article{l14,
  author={Fl{\"o}ry, Nikolaus and Ma, Ping and Salamin, Yannick and Emboras, Alexandros and Taniguchi, Takashi and Watanabe, Kenji and Leuthold, Juerg and Novotny, Lukas},
  title   = {{Waveguide-integrated van der Waals heterostructure photodetector at telecom wavelengths with high speed and high responsivity}},
  journal={Nature Nanotechnology},
  volume={15},
  number={2},
  pages={118--124},
  year={2020},
  note={\url {https://doi.org/10.1038/s41565-019-0602-z}}
}

@article{l17,
  author = {X. Huang and C. Liu and P. Zhou},
  title={{2D semiconductors for specific electronic applications: from device to system}},
  journal={npj 2D Materials and Applications},
  volume={6},
  number={1},
  pages={51},
  year={2022},
  note={\url {https://doi.org/10.1038/s41699-022-00327-3}}
}

@article{l18,
  author = {S. B. Cho and Y.-C. Chung},
  title={{Band engineering in a van der Waals heterostructure using a 2D polar material and a capping layer}},
  journal={Scientific Reports},
  volume={6},
  number={1},
  pages={27986},
  year={2016},
  note={\url {https://doi.org/10.1038/srep27986}}
}

@article{l19,
  title={{Superlattices based on van der Waals 2D materials}},
  author = {Y. K. Ryu and R. Frisenda and A. Castellanos-Gomez},
  journal={{Chemical Communications}},
  volume={55},
  number={77},
  pages={11498--11510},
  year={2019},
  publisher={Royal Society of Chemistry},
  note={\url {https://doi.org/10.1039/C9CC04919C}}
}

@article{l20,
  author = {K. Lin and G. Yao and J. Shao and Y. You and J. Qi and D. Yuan and Y. Wang and M. Wu and L. Kong and X. Zhang and others},
  title   = {{Nonlinear phase-matched van der Waals crystals integrated on optical fibres}},
  journal={Nature Materials},
  year    = {2026},
  note={\url {https://doi.org/10.1038/s41563-025-02461-x}}
}

@article{l21,
  author = {Q. Yan and Y. Weng and S. Wang and Z. Zhou and Y. Hu and Q. Li and J. Xue and Z. Feng and Z. Luo and R. Feng and others},
  title = {{Ambient Degradation Anisotropy and Mechanism of van der Waals Ferroelectric NbOI$_2$}},
  journal = {{ACS Applied Materials \& Interfaces}},
  volume = {16},
  number = {7},
  pages = {9051-9059},
  year = {2024},
  note={\url {https://doi.org/10.1021/acsami.3c18018}}
}

@article{l22,
  author={Abdelwahab, Ibrahim and Tilmann, Benjamin and Wu, Yaze and Giovanni, David and Verzhbitskiy, Ivan and Zhu, Menglong and Bert{\'e}, Rodrigo and Xuan, Fengyuan and Menezes, Leonardo de S and Eda, Goki and others},
  title   = {{Giant second-harmonic generation in ferroelectric NbOI$_2$}},
  journal={Nature Photonics},
  volume={16},
  number={9},
  pages={644--650},
  year={2022},
  note={\href {https://doi.org/10.1038/s41566-022-01021-y}
  {https://doi.org/10.1038/s41566-022-01021-y}}
}

@article{l23,
  author = {X. Ma and Q. Liu and D. Xu and Y. Zhu and S. Kim and Y. Cui and L. Zhong and M. Liu},
  title = {{Capillary-Force-Assisted Clean-Stamp Transfer of Two-Dimensional Materials}},
  journal = {Nano Letters},
  volume = {17},
  number = {11},
  pages = {6961-6967},
  year = {2017},
  note={\href {https://doi.org/10.1021/acs.nanolett.7b03449}
  {https://doi.org/10.1021/acs.nanolett.7b03449}}
}

@article{l25,
  author = {Y. Wang and K. D. Jöns and Z. Sun},
  title   = {{Integrated photon-pair sources with nonlinear optics}},
  journal = {Applied Physics Reviews},
  volume = {8},
  number = {1},
  pages = {011314},
  year = {2021},
  note={\href {https://doi.org/10.1063/5.0030258}
  {https://doi.org/10.1063/5.0030258}}
}

@article{Kwon08,
  author = {O. Kwon and Y.-W. Cho and Y.-H. Kim},
  title={{Single-mode coupling efficiencies of type-II spontaneous parametric down-conversion: Collinear, noncollinear, and beamlike phase matching}},
  journal={Physical Review A},
  year={2008},
  volume={78},
  pages={053825},
  note= {\url{https://doi.org/10.1103/PhysRevA.78.053825}}
}

@article{ref-transfer,
  title={{Transfer and beyond: emerging strategies and trends in two-dimensional material device fabrication}},
  author = {G. Huang and R. Chen and M. Chen and X. Chen and M. Jiang and Y. Xing and J. Wang and B. Liang and Q. Liu and X. Li and others},
  journal={Chemical Society Reviews},
  year={2026},
  publisher={Royal Society of Chemistry},
    note= {\url{https://doi.org/10.1039/D5CS00531K}}}

@article{joshi2026air,
  title={{Air-stable bright entangled photon-pair source from graphene-encapsulated van der Waals ferroelectric NbOI$_2$}},
  author = {M. Joshi and M. Jiang and Y. Xing and Y. Lu and J. Zhao and P. K. Lam and S. M. Assad and X. Ma and Y.-W. Cho},
  journal={arXiv preprint arXiv:2603.04082},
  year={2026},
  eprint={2603.04082},
   note= {\url{https://arxiv.org/abs/2603.04082}}
   }

@article{ma2021engineering,
  title={{Engineering photonic environments for two-dimensional materials}},
  author = {X. Ma and N. Youngblood and X. Liu and Y. Cheng and P. Cunha and K. Kudtarkar and X. Wang and S. Lan},
  journal={Nanophotonics},
  volume={10},
  number={3},
  pages={1031--1058},
  year={2021},
  publisher={De Gruyter},
   note= {\url{https://doi.org/10.1515/nanoph-2020-0524}}
}

@article{bai,
  title={{Photonic Crystal Defect Cavities Enable Air-Stable and Enhanced SHG from NbOCl$_2$}},
  author = {L. Bai and M. Jiang and L. Qu and P. Zhu and W. Wu and Q. Li and Z. Guo and X. Sun and Z. Huang and M. Xie and others},
  journal={Advanced Optical Materials},
  volume={13},
  number={30},
  pages={e01588},
  year={2025},
  publisher={Wiley Online Library},
  note= {\url{https://doi.org/10.1002/adom.202501588}}
}

@article{Br,
  title={{Extraordinary enhancement of nonlinear optical interaction in NbOBr$_2$ microcavities}},
  author = {W. Chen and S. Zhu and R. Duan and C. Wang and F. Wang and Y. Wu and M. Dai and J. Cui and S. H. Chae and Z. Li and others},
  journal={Advanced Materials},
  volume={36},
  number={26},
  pages={2400858},
  year={2024},
  publisher={Wiley Online Library},
  note= {\url{https://doi.org/10.1002/adma.202400858}}
}

@article{l28,
   title={{Counter-propagating entangled photon pairs from monolayer GaSe}},
  author = {Z. Lu and J. Janousek and S. M. Assad and S. Qiu and M. Joshi and Y. Hu and A. Y. Song and C. Wang and M. Suriyage and J. Zhao and others},
  journal={{Nature Communications}},
  volume={16},
  number={1},
  pages={9616},
  year={2025},
  note= {\url{https://doi.org/10.1038/s41467-025-64620-7}}
}

@article{l29,
  title={{Classical model of spontaneous parametric down-conversion}},
  author = {G. Kulkarni and J. Rioux and B. Braverman and M. V. Chekhova and R. W. Boyd},
  journal={{Physical Review Research}},
  volume={4},
  issue = {3},
  number={033098},
  year={2022},
    note= {\url{https://doi.org/10.1038/s41467-025-64620-7}}
}

@article{aletheia2025nonlinear,
  author = {W. Aletheia and M. Jiang and M. Cao and S. Huang and J. Lourembam and X. Ma and R. Duan and Z. Liu},
  title = {Nonlinear optical processes in 2D Cairo pentagonal palladium phosphide sulfide},
  journal={Nano Research},
  year={2025},
  publisher={Tsinghua University Press},
    note= {\url{https://doi.org/10.26599/NR.2026.94908387}}
}

@article{trovatello2025quasi,
  title={{Quasi-phase-matched up-and down-conversion in periodically poled layered semiconductors}},
  author = {C. Trovatello and C. Ferrante and B. Yang and J. Bajo and B. Braun and Z. H. Peng and X. Xu and P. K. Jenke and A. Ye and M. Delor and others},
  journal={Nature Photonics},
  volume={19},
  number={3},
  pages={291--299},
  year={2025},
  publisher={Nature Publishing Group UK London},
  note= {\url{https://doi.org/10.1038/s41566-024-01602-z}}
}

@article{feng24,
  author = {J. Feng and Y.-K. Wu and R. Duan and J. Wang and W. Chen and J. Qin and Z. Liu and G.-C. Guo and X.-F. Ren and C.-W. Qiu},
  title={{Polarization-entangled photon-pair source with van der Waals 3R-WS$_2$ crystal}},
  journal = {eLight},
  volume  = {4},
  number={1},
  pages   = {16},
  year    = {2024},
  note = {\url{https://doi.org/10.1186/s43593-024-00074-6}}
}

@article{Hong2021,
  author = {S. Hong and J. ur Rehman and Y. S. Kim and Y. W. Cho and S. W. Lee and H. Jung and S. Moon and S. W. Han and H. T. Lim},
  title={{Quantum enhanced multiple-phase estimation with multi-mode N00N states}},
  journal={Nature Communications},
  volume={12},
  number={1},
  pages={5211},
  year={2021},
  publisher={Nature Publishing Group UK London},
  note = {\url {https://doi.org/10.1038/s41467-021-25451-4}}
}

@article{Hong2022,
  title={{Practical Sensitivity Bound for Multiple Phase Estimation with Multi-Mode N 00 N N00N States}},
  author={Hong, Seongjin and Rehman, Junaid ur and Kim, Yong-Su and Cho, Young-Wook and Lee, Seung-Woo and Lee, Su-Yong and Lim, Hyang-Tag},
  journal={Laser \& Photonics Reviews},
  volume={16},
  number={9},
  pages={2100682},
  year={2022},
  publisher={Wiley Online Library},
  note = {\url {https://doi.org/10.1002/lpor.202100682}}
}

@article{Xuan2024,
  author = {F. Xuan and M. Lai and Y. Wu and S. Y. Quek},
  title={{Exciton-Enhanced Spontaneous Parametric Down-Conversion in Two-Dimensional Crystals}},
  journal={Physical Review Letters},
  volume={132},
  number={24},
  pages={246902},
  year={2024},
  publisher={American Physical Society},
  note = {\url {https://doi.org/10.1103/PhysRevLett.132.246902}}
}

@article{HBT1956,
  author = {R. Hanbury Brown and R. Q. Twiss},
  title={{Correlation between Photons in Two Coherent Beams of Light}},
  journal={Nature},
  volume={177},
  number={4497},
  pages={27--29},
  year={1956},
  publisher={Nature Publishing Group UK London},
  note = {\url {https://doi.org/10.1038/177027a0}}
}

@article{Guerreiro2013,
  author = {T. Guerreiro and A. Martin and B. Sanguinetti and R. Thew and H. Zbinden},
  title={{High efficiency coupling of photon pairs in practice}},
  journal={Optics Express},
  volume={21},
  number={23},
  pages={27641--27651},
  year={2013},
  publisher={Optica Publishing Group},
  note = {\url {https://doi.org/10.1364/OE.21.027641}}
}

@article{Schwaller2022,
  author = {N. Schwaller and others},
  title={{Optimizing the correlated-mode coupling efficiency of SPDC photon pairs}},
  journal={Physical Review A},
  volume={106},
  pages={043719},
  year={2022},
  publisher={American Physical Society},
  note = {\url {https://doi.org/10.1103/PhysRevA.106.043719}}
}

@article{Okoth2019,
  author = {C. Okoth and A. Cavanna and A. Di Falco},
  title={{Microscale Generation of Entangled Photons without Momentum Conservation}},
  journal={Physical Review Letters},
  volume={123},
  number={26},
  pages={263602},
  year={2019},
  publisher={American Physical Society},
  note = {\url {https://doi.org/10.1103/PhysRevLett.123.263602} }
}

@article{Ling2008,
  author = {A. Ling and A. Lamas-Linares and C. Kurtsiefer},
  title={{Absolute emission rates of spontaneous parametric down-conversion into single transverse Gaussian modes}},
  journal={Physical Review A},
  volume={77},
  number={4},
  pages={043834},
  year={2008},
  publisher={American Physical Society},
  note = {\url {https://doi.org/10.1103/PhysRevA.77.043834}} 
}

@article{Wang2016,
  author = {Wang, Xi-Lin and Chen, Luo-Kan and Li, Wei and Huang, H-L and Liu, Chang and Chen, Chao and Luo, Y-H and Su, Z-E and Wu, Dian and Li, Z-D and others},
  title={{Experimental Ten-Photon Entanglement}},
  journal={Physical Review Letters},
  volume={117},
  number={21},
  pages={210502},
  year={2016},
  publisher={American Physical Society},
  note = {\url{https://doi.org/10.1103/PhysRevLett.117.210502}
}}

@article{Shalm2015,
  author = {L. K. Shalm and E. Meyer-Scott and B. G. Christensen and P. Bierhorst and M. A. Wayne and M. J. Stevens and T. Gerrits and S. Glancy and D. R. Hamel and M. S. Allman and others},
  title={{Strong Loophole-Free Test of Local Realism}},
  journal={Physical Review Letters},
  volume={115},
  number={25},
  pages={250402},
  year={2015},
  publisher={American Physical Society},
  note = {\url{https://doi.org/10.1103/PhysRevLett.115.250402}}
}

@article{Walborn2010,
  author = {Walborn, Stephen P and Monken, CH and P{\'a}dua, S and Ribeiro, PH Souto},
  title={{Spatial correlations in parametric down-conversion}},
  journal={Physics Reports},
  volume={495},
  number={4--5},
  pages={87--139},
  year={2010},
  publisher={Elsevier},
  note = {\url {https://doi.org/10.1016/j.physrep.2010.06.003}}
}

@article{Choi2020,
  author = {Y.-H. Choi and S. Hong and T. Pramanik and H.-T. Lim and Y.-S. Kim and H. Jung and S.-W. Han and S. Moon and Y.-S. Kim},
  title={{Demonstration of simultaneous quantum steering by multiple observers via sequential weak measurements}},
  journal={Optica},
  volume={7},
  number={6},
  pages={675--679},
  year={2020},
  publisher={Optica Publishing Group},
  note = {\url{https://doi.org/10.1364/OPTICA.394667}}
}

@article{Cho2019,
  author = {Y.-W. Cho and Y. Kim and Y.-H. Choi and Y.-S. Kim and S.-W. Han and S.-Y. Lee and S. Moon and Y.-H. Kim},
  title={{Emergence of the geometric phase from quantum measurement back-action}},
  journal={Nature Physics},
  volume={15},
  number={7},
  pages={665--670},
  year={2019},
  publisher={Nature Publishing Group},
  note = {\url {https://doi.org/10.1038/s41567-019-0482-z}}
}
\end{document}